\documentclass[%
 reprint,
 amsmath,amssymb,
 aps,
]{revtex4-2}

\usepackage{graphicx}
\usepackage{dcolumn}
\usepackage{bm}

\begin{document}

\title{Non-equilibrium relaxation process of complex systems and its statistical physical properties}

\author{Zhifu Huang}
\email{zfhuang@hqu.edu.cn}

\author{Yuqing Wang}

\affiliation{College of Information Science and Engineering, Huaqiao University, Xiamen 361021, People's Republic of China}

\date{\today}

\begin{abstract}
We construct a model that evolved from non-equilibrium relaxation process. The model needs only two basic coefficients which can be directly obtained from real data, including self-similar coefficient and non-equilibrium coefficient. Besides that, the model can well represent the phenomena of anomalous diffusion, non-Gaussian distribution and so on. Importantly, the model use global iterative method that can exhibit statistical physical properties in multiple time scales, as well as exhibiting long-term memory properties. It may be used as a basic model for the research of various anomalous phenomena in complex systems.
\end{abstract}

\pacs{05.65.+b; 05.45.Ac; 02.50.Cw}

\maketitle

There is no absolute stillness, only relative movement. Researching the changing process of matter can often be simplified to analyzing the trajectory of particles. Because affected by too many factors in complex systems, the trajectory of particles often cannot be deterministically described. In this case, it becomes important to study the statistical physical properties of particle motion \cite{ein}.

In various fields of science, there are complex system problems composed of a large number of individuals, including physics, chemistry, biology, earthquakes, brain science, economics, society and so on. In these different scientific fields, it comes to similar phenomena, such as long-range interaction \cite{fre10, long17}, long-term memory \cite{lm13, lm14}, anomalous diffusion \cite{rka00, ad15}, fat-tail distribution \cite{ba05, fd19} and so on. Previous researchers often used probability and statistical methods to study complex system problems. The most important theorem in the field of probability statistics is central limit theorem, its basic assumption includes that a single event must be a nearly independent event, and its higher moments cannot diverge. For the problem of higher moment divergence, there have been many research results in different fields in the past, such as  L\'{e}vy process \cite{mfsh00}, fractional stable L\'{e}vy motions \cite{rmjk00}, truncated L\'{e}vy process \cite{rnmh00} and so on.
Moreover, the complex systems mentioned above often exhibit the effect of long-term memory. However, none of these studies have targeted the long-term memory problems. In other words, non-nearly independent events that may exist in complex systems have not been researched in depth.  Therefore, the long-term memory is an important problem that must be analyzed.

It is well known, due to many influence factors, the particles in a complex system are in constant movement, and may continuously transform between non-equilibrium and equilibrium states. Previous researches tell us that after a non-equilibrium relaxation time, the particle will be in equilibrium state. However, how particles gradually transform from a non-equilibrium state to an equilibrium state during the relaxation process has not been discussed in detail in the past. This letter will start from the basic concept of relaxation time, and research the physical statistical properties of the transition from non-equilibrium state to equilibrium state in complex systems, as well as the long-term memory properties of complex systems during non-equilibrium relaxation process.

Aiming to study the most basic properties, we briefly discuss the behavior of a particle in complex system in one dimension. In experimental systems, the time interval of any process is finite. When ensuring the position, the displacement during the time interval $\Delta t$ can be defined as $x(t,\Delta t)=s(t+\Delta t)-s(t) $, where $s(t)$ is the position at time $t$, and the corresponding speed can be given by $v(t,\Delta t)=x(t,\Delta t)/\Delta t $. Note that the displacement and velocity mentioned here are not instantaneous values. They depend not only on the time, but also on the time interval. Accordingly, the displacement and velocity in different time intervals may need to be analyzed.

The displacement of the particle during anomalous diffusion is a typical time interval dependent series. These anomalous diffusions can be characterized by one-dimensional mean-square displacement as $ \langle x^{2}(t,\Delta t)\rangle_{t}\propto\Delta t^{\alpha}$, where $\langle...\rangle_{t}$ means averaging the values of the quantities at different time $t$, and $\alpha$ is called the anomalous diffusion coefficient.
The cases of $\alpha<1$ and $\alpha>1$ correspond to the subdiffusion and superdiffusion, respectively.  $\alpha =1$ represents the normal diffusion, which is the result of the Brownian motion.
On the basis of the definition of velocity, one can obtain the mean-square velocity during the time interval as $ \langle v^{2}(t,\Delta t)\rangle_{t}\propto\Delta t^{\alpha-2}$.
It can be seen that when $0<\alpha<2$, the mean-square displacement is convergent but the mean-square velocity is divergent at the $\Delta t\rightarrow0$ limit.
Since the mean-square velocity is divergent at the $\Delta t\rightarrow0$ limit, the instantaneous velocity can not be strictly measured.
 In the situation that a moving object may require an infinite force to change its speed or direction discontinuously, neither the instantaneous force nor the average force can be defined. According to the law of mechanics, we can define an available force as
\begin{equation}
F(t,\Delta t)
\label{af}=m\frac{v(t+\Delta t,\Delta t)-v(t,\Delta t)}{\Delta t},
\end{equation}
where $m$ is the mass of the moving object and we set $m=1$ for the sake of convenience. It is worthwhile to mention that in different time intervals, the available force (AF) has different values.

The research field of statistical physics is very extensive, and many different complex systems exhibit similar statistical physical properties. However, different complex systems may exhibit different specific interactions. Hence, in the following analysis, the interaction relationship of the complex system is no longer the focus of our research.
On the other hand, for the characteristics of the AF obtained above, it must be noted that the value of the AF is different in different time intervals. One can study the statistical physical properties of complex systems from the relationship of AF in different time intervals.
It is well known that complex systems tend to have self-similarity properties, and statistical physics quantities at different time scales have similarities \cite{ss82}. For example, for Brownian motion, one can get the standard deviation of AF in different time interval with $ \sigma_{F}(\Delta t)=2\sqrt{2}\sigma_{F}(2\Delta t)$, where $\sigma_{F}(\Delta t)=\sqrt{\langle F^{2}(t,\Delta t)\rangle_{t}}$ means the standard deviation of the AF in the time interval $\Delta t$. As a result, the model we introduced also attempts to consider the relationship of the standard deviation of AF in different time scales, and use coefficient $C$ to represent self-similarity.

In a complex system, the particles will be in equilibrium state after the relaxation time. The statistical physical theory of equilibrium physics has been relatively perfect. For this reason, we can only pay attention to the problem that the time interval of particle motion is below the relaxation time. Due to influenced by many factors in a complex system, the trajectory of the same particle undergoing relaxation process may be different, so we can consider researching the statistical properties of the trajectory of the same particle undergoing multiple relaxation processes. For example, we can calculate the standard deviation of the AF at the same time $t$ and time interval $\Delta t$ in multiple relaxation processes.
Below the relaxation time, because the complex system is in non-equilibrium state, the standard deviation of the AF at different time $t$ may be different.
The simplest case can be that the standard deviation at the first half of the relaxation time is different from the second half of the relaxation time.
Therefore, we can introduce a non-equilibrium coefficient to describe the non-equilibrium properties of a complex system. In the past researches, a step-by-step iterative approach was often used to build models. For example, the Markov process uses a transition matrix to evolve the state of the system from the previous step to the next step. In the Markov process, since the state of the system is only derived from the state of the previous step, it cannot truly reflect the long-term memory of the system.
On the other hand, since the properties of complex systems in different scales are self-similar, we can introduce the idea of global iteration in different time scales. In addition, in order to ensure that the model is inversely symmetric in time, our iterative relationship is as follows
\begin{equation}
\label{map}
D_{F}(t,\Delta t)=\left\{
\begin{aligned}
&CD_{F}(2t,2\Delta t)  , & 0\leq t<\tau/2 \\
&C\lambda D_{F}[2(\tau-t),2\Delta t]  , & \tau/2\leq t<\tau ,
\end{aligned}
\right.
\end{equation}
where $\tau$ is relaxation time, $C$ is self-similar coefficient and $\lambda$ is non-equilibrium coefficient.
Note that $D_{F}(t,\Delta t)$ refers to the standard deviation of the $F(t,\Delta t)$ obtained in the multiple simulations for the same time $t$ and time interval $\Delta t$.
Since there is no correlation between different relaxation processes, we can assume that for the same time $t$ and time interval $\Delta t$, the AFs obtained from multiple simulations satisfy the normal distribution, that is $F(t,\Delta t)=D_{F}(t,\Delta t)\varepsilon$, where $\varepsilon$ is a normally distributed random number with an average value of zero and a standard deviation of 1.
According to the anomalous diffusion relationship, it can be obtained that the relationship between the variance of the AF and the time interval is  $ \langle F^{2}(t,\Delta t)\rangle_{t}\propto\Delta t^{\alpha-4}$.
Combined with the model definition ~(\ref{map}), the relationship satisfied by the anomalous diffusion coefficient, self-similar coefficient and non-equilibrium coefficient can be obtained as $\alpha=5-\ln[C^{2}(1+\lambda^{2})]/\ln2$.
It can be seen from the equation that the model can satisfy the anomalous diffusion relationship. Therefore, compare with the anomalous diffusion relationship, one can obtain parameter values of the model from real data without knowing the interaction of system.

In general, $C$ and $\lambda$ can also be related to the time interval, as $C=C(\Delta t)$ and $\lambda=\lambda(\Delta t)$. On this basis, the anomalous diffusion coefficient also changes with time intervals $\Delta t$.
Previous researches have also found that the anomalous diffusion coefficient changes with time intervals \cite{pu11}. Therefore, the relationship of the variance of the AF in different time scales can be obtained as $\sigma^{2}_{F}(\Delta t)=C^{2}(\Delta t)[1+\lambda^{2}(\Delta t)]\sigma^{2}_{F}(2\Delta t)/2$.

When the system is in non-equilibrium state, the distribution function of the physical quantity may no be normal distribution. In order to describe the deviation from the normal distribution, here we introduce kurtosis to measure it as
\begin{equation}
\label{fd}
\eta_{F}(\Delta t)=\frac{<F^{2}(t,\Delta t)>_{t}}{<|F(t,\Delta t)|>^{2}_{t}},
\end{equation}
where $\eta_{F}(\Delta t)$ is the kurtosis of AF in time interval $\Delta t$.
For a normal distribution, one can easily calculate that the corresponding kurtosis value is equal to half of $\pi$. Therefore, when the time interval of movement is greater than the relaxation time, the properties of the equilibrium state can be obtained, and the kurtosis value is half of $\pi$. On the other hand, when the time interval of movement is less than the relaxation time, the kurtosis value in non-equilibrium state may be greater than half of $\pi$.

In order to compare with the non-equilibrium relaxation process of the actual system, we selected foreign exchange transaction data as the research object.
The statistical properties of currency market fluctuations are important for modeling and understanding complex systems \cite{aab01}.
We study financial markets in this letter, namely, foreign exchange markets focused on the Euro vs U.S. Dollar exchange rate (EUR/USD) to compare with the non-equilibrium relaxation process of actual systems. We acknowledge metaquotes.net for providing all currency exchange data. We have selected the EUR/USD closing exchange prices in time intervals of 1 minute.
For the sake of convenience, we can define a basic quantity \cite{mrsh00}: the position $s(t)$ is the logarithmic of the exchange price $s(t)=\ln[price(t)]$. After that, the corresponding displacement, velocity and available force can be obtained. It must be mentioned that the displacement in our work is usually referred as return \cite{mrsh00}.

\begin{figure}
\includegraphics[width=8truecm]{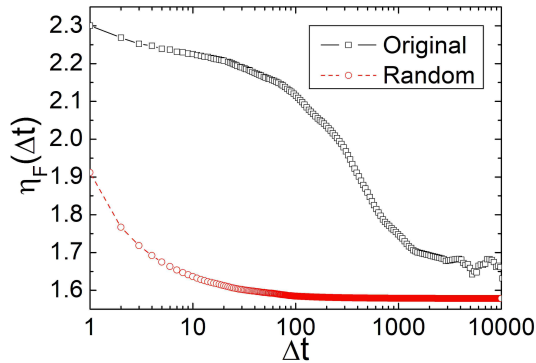}
\caption{\label{fig:fdpic}
The curve of kurtosis versus time interval.}
\end{figure}

Figure~\ref{fig:fdpic} shows the evolution of kurtosis over time interval. It can be clearly seen that when the time interval is large enough, since the motion behavior is close to the equilibrium state, the value of kurtosis is close to the normal distribution. On the other hand, when the time interval is relatively small, we can clearly see that the kurtosis deviates from the equilibrium state.
Furthermore, we can consider the influence of long-term memory on the evolution of kurtosis with respect to time intervals.
If we rearrange the velocity series randomly to eliminate the long-term memory, the distribution of displacement will reduce to the normal distribution. We can first randomize the velocity sequence at a time interval of 1 minute, and then add them together to calculate the random displacement at a certain time interval.
 In this case, the corresponding displacement can be calculated by
\begin{equation}
\label{ran}
x_{ran}(t,\Delta t)=\sum^{\Delta t/\Delta t_{0}}_{1}v_{ran}(t,\Delta t_{0})\Delta t_{0}.
\end{equation}
Where the time interval of $\Delta t_{0}$ is equal to 1 minute, and $v_{ran}(t,\Delta t_{0})$ is the discrete value that is randomly chosen from the velocity series.
After the displacement $x_{ran}(t,\Delta t)$ is obtained, we can calculate the corresponding AF. It can be seen in figure ~\ref{fig:fdpic} that the relationship between the evolution of kurtosis and the time interval after randomization is different from that of the case without disruption.
In previous work, such as L\'{e}vy motions and fractional stable L\'{e}vy motions, truncated L\'{e}vy process, the time series evolution is one-step or multi-step iterations, it cannot present the characteristics of long-term memory. As a result, the presence or absence of randomization does not affect the evolution of kurtosis with time intervals.
On the contrary, our work consider the evolution relationship of kurtosis with time interval during non-equilibrium relaxation process. Then, the long-term memory effect of complex systems will be characterized.

\begin{figure}
\includegraphics[width=8truecm]{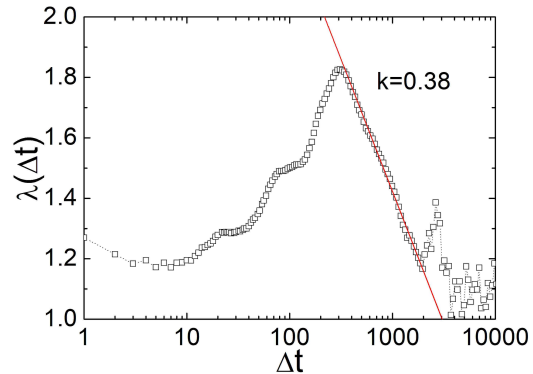}
\caption{\label{fig:fdbipic}
The curve of non-equilibrium coefficient versus time interval. k is a dimensionless constant describing the relaxation relationship of the non-equilibrium coefficient with respect to the time interval.}
\end{figure}

According to the iterative relationship of the model ~(\ref{map}), one can easily get the relationship between the kurtosis of the AF and the time interval as
\begin{equation}
\label{fdbi}
\eta_{F}(\Delta t)=\frac{2[1+\lambda^{2}(\Delta t)]}{[1+\lambda(\Delta t)]^{2}}\eta_{F}(2\Delta t).
\end{equation}
It can be clearly seen that the relationship between the kurtosis of the AF and the time interval is only related to the non-equilibrium coefficient. Therefore, based on the real data, the non-equilibrium coefficient can be directly obtained.
It can be seen in figure~\ref{fig:fdbipic} that, for foreign exchange data, when the time interval is large enough, the non-equilibrium coefficient is close to 1, which is consistent with the common sense that when the time interval is relatively large, the movement is close to Brownian motion. On the other hand, one can also see that the non-equilibrium coefficient reaches its maximum value at a time interval of about 300 minutes.
As the time interval becomes larger, the system may gradually relax, and the non-equilibrium coefficient is getting closer to 1. In addition, as the time interval becomes larger, the change rate of the non-equilibrium coefficient with the time interval becomes smaller and smaller. Therefore, considering only the simplest first order term, we can give the relaxation relationship of the non-equilibrium coefficient with respect to the time interval as follow
\begin{equation}
\label{k}
\frac{d\lambda(\Delta t)}{d\Delta t}=-\frac{k}{\Delta t}.
\end{equation}
Where $k$ is a dimensionless constant greater than zero, and we can call it relaxation coefficient. In figure~\ref{fig:fdbipic}, one can clearly see that within a certain time interval, the above relaxation relationship satisfies the data very well, and the value of k is 0.38.
It must be emphasized that the relaxation coefficient obtained here is a dimensionless parameter, which can describe the evolution relationship of the non-equilibrium properties with time in the relaxation process, and may be used to describe the universal property of the non-equilibrium relaxation process in the future. The ideal relaxation time may be infinite, but under the conditions of satisfying the above relaxation relationship, we can calculate the time interval when the non-equilibrium coefficient is equal to 1. In the data of this work, we get the corresponding time interval of 3043 minutes.

\begin{figure}
\includegraphics[width=8truecm]{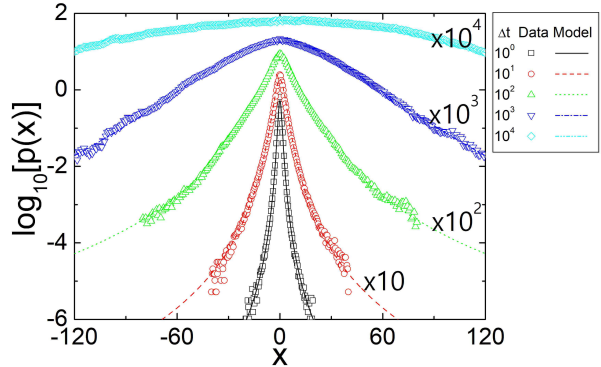}
\caption{\label{fig:pdf}
The probability distribution function of displacement in different time interval.}
\end{figure}

In addition, when the time interval exceeds the relaxation time, the system is in equilibrium state. Therefore, we can take the case when the time interval is relatively large, the movement of the particle is the behavior of Brownian motion, and then globally iterating the AF time series. In this work, we choose the time interval of $2^{15}$ minutes as the equilibrium state, and the standard deviation of the AF under this time interval does not change with time, as $D_{F}(t,\tau)=\sigma_{F}(\tau)$.
After that, we can get the standard deviation series of the AF at each time interval, and then we can simulate the time series of the AF. With the time series of AF, we can get the time series of the corresponding velocity and the position based on the definition of motion. In order to make an intuitive comparison between the foreign exchange data and the model, we make the 1 minute of the foreign exchange data equivalent to the model $\Delta t=1$. For example, the position value of the foreign exchange data is divided by the standard deviation of displacement in 1 minute, and the position value of the model is also divided by the standard deviation of displacement under $\Delta t=1$.

After the time series of the model has been constructed, we can compare the probability distribution functions of displacement in different time interval obtained from the model and the real data. In figure~\ref{fig:pdf}, one can see that in different time scales, the model and the real data can be in good agreement. What needs to be emphasized here is that in the process of constructing this model, without knowing the interaction of the system, we still reconstruct the non-equilibrium statistical physical properties of the system in different time scales. Compared with the previous work, the fat-tail distribution can be obtained from the high-order moment divergence, we only consider the long-term memory here, and the fat-tail distribution has also been obtained.
It can be seen from figure~\ref{fig:fdpic} that only considering the properties of the high-order moment divergence cannot describe the non-equilibrium relaxation behavior of the real system. In addition, the past financial models generally only give statistical properties in one time scale. Our model can reconstruct the statistical physical properties in multiple time scales, and given the concrete model of the multi-scale properties of financial price changes revealed in the past work \cite{gh96}. Therefore, this model may be used as the basic model of the financial price system, or it may be used as the basic model of other complex systems when the interaction relationships cannot be known.

Starting from the non-equilibrium relaxation process, we construct a model that the coefficients can be directly obtained from the real data.
The model can well represent the important statistical physical properties such as anomalous diffusion, non-Gaussian distribution and so on that are common in real complex systems. It is important to mention that, unlike previous work, the model performs a global iteration which can exhibit statistical physical properties in different time scales, while also exhibiting long-term memory phenomena. It is particularly emphasized that the model does not need to know the interaction relationship of the system.
This model may open a new perspective for complex system problems in many scientific fields.

\begin{acknowledgments}
Project supported by the National Natural Science Foundation (No. 11775084,11305064).
\end{acknowledgments}

\end{document}